\documentclass[a4paper,11pt]{article}
 \pdfoutput=1
 \usepackage{lscape}
 \usepackage[all]{xy} 
 \usepackage{longtable} 
 \usepackage{jheppub} 
 \usepackage{graphicx}
 \usepackage{amssymb}
 \usepackage{amsmath}
\usepackage{mathtools}
\usepackage{listings}
\usepackage{dcolumn}
\usepackage{bm}
\usepackage{color}
\usepackage{multirow}
\usepackage{upquote} 
\usepackage{subcaption}

\allowdisplaybreaks

\newcommand{\lsim}{{\;\raise0.3ex\hbox{$<$\kern-0.75em\raise-1.1ex\hbox{$\sim$}
}\;}}
\newcommand{\gsim}{{\;\raise0.3ex\hbox{$>$\kern-0.75em\raise-1.1ex\hbox{$\sim$}
}\;}}
\newcommand{\beq}{\begin{equation}}
\newcommand{\eeq}{\end{equation}}
\newcommand{\bea}{\begin{eqnarray}}
\newcommand{\eea}{\end{eqnarray}}
\mathchardef\minus="002D

\newcommand{\be} {\begin{equation}}
\newcommand{\ee} {\end{equation}}
\newcommand{\ba} {\begin{eqnarray}}
\newcommand{\ea} {\end{eqnarray}}

\newcommand{\no} {\nonumber}
\newcommand{\cL} {\mathcal L}
\newcommand{\cA} {\mathcal A}

\title{\boldmath Adding Pseudo-Observables to the Four-Lepton
  Experimentalist's Toolbox}

\author[a]{James S. Gainer,}
\author[b]{Mart\'in Gonz\'alez-Alonso,}
\author[c, d]{Admir Greljo,}
\author[d]{Senad Isakovi\' c,}
\author[e]{Gino Isidori,}
\author[f]{Andrey Korytov,}
\author[g]{Joseph Lykken,}
\author[h]{David Marzocca,}
\author[f]{Konstantin T. Matchev,}  
\author[i]{Predrag Milenovi\' c,}
\author[f]{Guenakh Mitselmakher,}
\author[j]{Stephen Mrenna,} 
\author[k]{Myeonghun Park,}
\author[l]{Aurelijus Rinkevicius,}
\author[d]{and Nud\v zeim Selimovi\' c}

 \affiliation[a]{Dept. of Physics and Astronomy, University of Hawaii,
Honolulu, HI 96822, USA}
\affiliation[b]{Theoretical Physics Department, CERN, 1211 Geneva 23,
Switzerland}
 \affiliation[c]{PRISMA Cluster of Excellence and Mainz Institute for
Theoretical Physics, Johannes Gutenberg Universit\"{a}t
Mainz, 55099 Mainz, Germany}
\affiliation[d]{Faculty of Science, University of Sarajevo, 
Zmaja od Bosne 33-35, 71000 Sarajevo, Bosnia and
Herzegovina}
\affiliation[e]{Physik-Institut, Universit\"at Z\"urich, CH-8057
Z\"urich, Switzerland}
  \affiliation[f]{Physics Department, University of Florida,
  Gainesville, FL 32611, USA}
\affiliation[g]{Theoretical Physics Department, Fermilab, Batavia, IL
  60510, USA}
\affiliation[h]{INFN, Sezione di Trieste, SISSA, Via Bonomea 265, 34136,
Trieste, Italy}
\affiliation[i]{Experimental Physics Department, CERN, 1211 Geneva 23,
Switzerland}
\affiliation[j]{SSE Group, Computing Division, Fermilab, Batavia, IL
  60510, USA} 
\affiliation[k]{Institute of Convergence Fundamental Studies and School of
Liberal Arts,
Seoultech, Seoul 01811, Korea} 
\affiliation[l]{Department of Physics, Cornell University, Ithaca, NY 14850,
USA}

\abstract{
The ``golden'' channel, in which the newly-discovered Higgs boson
decays to four leptons by means of intermediate vector bosons, is important for determining the properties of the Higgs boson and for
searching for subtle new physics effects.  Different approaches exist
for parametrizing the relevant Higgs couplings in this channel; here
we relate the use of pseudo-observables to methods based on specifying
the most general amplitude or Lagrangian terms for the $HVV$
interactions.  We also provide projections for sensitivity in this
channel in several novel scenarios, illustrating the use of pseudo-observables, and analyze the role of kinematic distributions and (ratios of) rates in such $H\to4\ell$ studies.
}

\date{August 1, 2018}
\preprint{
\begin{flushright} 
CERN-TH-2018-178
\\
FERMILAB-PUB-18-332-CD
\\
UH-511-1255-18
\end{flushright} 
}

\begin{document}
\maketitle
\flushbottom

\section{Introduction}
\label{sec:introduction}
The discovery of a Standard Model (SM)-like Higgs boson (Higgs) at the CERN
Large Hadron Collider (LHC)~\cite{Aad:2012tfa, Chatrchyan:2012xdj}
represents both the triumphant coda to one era,
in which the SM was developed, tested, and confirmed, and the
dawn of an exciting new era, in which physics beyond the SM (BSM) is
probed, and eventually discovered.  The ``golden'' channel, in which the Higgs decays to vector bosons (generally, $Z$ bosons) that consequently decay to leptons, has played an important role in both the
discovery of the Higgs and in the subsequent measurement of its
couplings.  It has therefore played an important role in both
experimental~\cite{
    Aad:2012tfa, Chatrchyan:2012xdj, Aad:2011qi, Aad:2011uq,
    ATLAS:2012ae, ATLAS:2012ac, Chatrchyan:2012tx, Chatrchyan:2012dg,
    Aad:2012an, CMS:2012bw, ATLAS:2012oga, Chatrchyan:2012rva,
    Chatrchyan:2012jja, Chatrchyan:2013lba, Chatrchyan:2013yoa,
    Aad:2013wqa, Aad:2013xqa, Chatrchyan:2013mxa, Khachatryan:2014iha,
    Aad:2014aba, Aad:2014tca, Khachatryan:2014kca, Khachatryan:2014jba,
    Aad:2015xua, Aad:2015zhl, Khachatryan:2015cwa, Aad:2015lha,
    Aad:2015sva, Khachatryan:2015wka, Aad:2015mxa, Aad:2015kna,
    Khachatryan:2015mma, Khachatryan:2015yvw, Khachatryan:2016tnr,
    Khachatryan:2016ctc, Khachatryan:2016vau, Aaboud:2016okv,
    Sirunyan:2017exp, Sirunyan:2017tqd, Aaboud:2017vzb, Sirunyan:2018qlb}
and theoretical~\cite{DellAquila:1985mtb, Nelson:1986ki, Kniehl:1990yb,
    Soni:1993jc, Chang:1993jy, Barger:1993wt, Arens:1994wd, Choi:2002jk,
    Allanach:2002gn, Buszello:2002uu, Schalla:2004ura, Godbole:2007cn,
    Kovalchuk:2008zz, Keung:2008ve, Antipin:2008hj, Cao:2009ah,
    Gao:2010qx, DeRujula:2010ys, Englert:2010ud, Matsuzaki:2011ch,
    DeSanctis:2011yc, Logan:2011ey, Gainer:2011xz, Low:2011kp,
    Englert:2012ct, Campbell:2012cz, Campbell:2012ct, Kauer:2012hd,
    Kniehl:2012rz, Moffat:2012pb, Coleppa:2012eh, Bolognesi:2012mm,
    Boughezal:2012tz, Stolarski:2012ps, Cea:2012ud, Kumar:2012ba,
    Geng:2012hy, Avery:2012um, Masso:2012eq, Chen:2012jy, Modak:2013sb,
    Kanemura:2013mc, Gainer:2013rxa, Isidori:2013cla, Frank:2013gca,
    Grinstein:2013vsa, Caola:2013yja, Banerjee:2013apa, Sun:2013yra,
    Anderson:2013afp, Chen:2013waa, Buchalla:2013mpa, Chen:2013ejz,
    Campbell:2013una, Curtin:2013fra, Chen:2014pia,
    Gonzalez-Alonso:2014rla, Gainer:2014hha,
    Chen:2014gka, Falkowski:2014ffa, Beneke:2014sba,
    Gao:2014xlv, Modak:2014zca,
    Chen:2014hqs, Modak:2014ywa, Beneke:2014vqa,
    Curtin:2014cca,
    Belyaev:2015xwa, Englert:2015zra,
    Chen:2015iha, Bhattacherjee:2015xra, Furui:2015foa,
    Gonzalez-Alonso:2015bha,
    Zagoskin:2015sca, Banerjee:2015bla, Zhou:2015wra,
    Chen:2015rha, Curtin:2015fna, Davoudiasl:2015bua,
    Djouadi:2015aba, Delgado:2016arn, Dwivedi:2016xwm, Caola:2016trd,
    Chen:2016ofc, Ginzburg:2016pbh, Grojean:2016pvc, deFlorian:2016spz,
    Zagoskin:2017bzx, Chang:2017kmk, Vega:2017gkk, Brehmer:2017lrt,
    Zagoskin:2018wdo, Brehmer:2018kdj, Brehmer:2018eca}
studies of the Higgs boson.

Measuring the couplings of the Higgs boson, and thus searching for
the small deviations from SM predictions that may be the hallmark of
BSM physics, requires a parameterization of these couplings.
We can view this parameterization as being determined by:
\begin{itemize}
    \item \textbf{Particles}: First we must specify the spectrum of
          particles that appear, either in the final state or as intermediate
          resonances.
    \item \textbf{Symmetries}: Next, we must determine the symmetries
          that the underlying theory obeys, either exactly or approximately.
    \item \textbf{Interactions}:	Once the particles and symmetries have been
          specified, we can enumerate the possible interactions, all of
          which we expect to be present at some level.  However, in general,
          there are additional considerations, such as the dimensionality of
          operators in an effective field theory (EFT) Lagrangian, which may
          allow us to consider only a finite number of terms.
\end{itemize}

In this paper we will consider parameterizations relevant to
experimental studies of the four-lepton final state at the LHC.  These
parametrizations are based on the choices of
\begin{enumerate}
    \item \textbf{Particles}: We consider only the SM degrees of freedom,  with an additional (optional) heavy $Z^\prime$ boson, which is introduced as a BSM example of a concrete BSM scenario that allows for different couplings to electrons and muons. 
          In particular, we do not consider the effects of light 
          $Z^\prime$s~\cite{Aad:2015sva, Curtin:2013fra, Gonzalez-Alonso:2014rla,
          Falkowski:2014ffa, Gao:2014xlv, Modak:2014ywa, Curtin:2014cca,
          Curtin:2015fna, Davoudiasl:2015bua}, dark matter, or other
          invisible or unidentified particles in the final state~\cite{Wise:1981ry, Shrock:1982kd, Griest:1987qv,
          Romao:1992zx, Gunion:1993jf, Choudhury:1993hv,
          Frederiksen:1994me, Djouadi:1996mj, Martin:1999qf}.
    \item \textbf{Symmetries}: There have been many studies of Higgs
          couplings that make the (highly theoretically motivated) assumption
          that the Higgs is part of an $SU(2)$ doublet and the only allowed
          operators are part of the SM $SU(3) \times SU(2) \times U(1)$ gauge
          symmetry.  We, however, make fewer theoretical assumptions as our
          aim is to facilitate relatively unbiased experimental measurements.
          If, as expected, potential interactions, which violate the above
          assumptions, are absent, we have obtained additional
          experimental support for our pre-existing viewpoint.
          Thus, we will assume only Lorentz invariance and QED and QCD gauge
          invariance in specifying possible interactions.
    \item \textbf{Interactions}: Having made these choices of particles
          and symmetries, the allowed interactions are restricted.  We will
          generally truncate the (infinite) list of possible interactions by
          considering operators only up to some finite mass dimension or
          (equivalently) amplitude structures only up to a specified power of
          momentum.
\end{enumerate}

Experimental attempts to measure the tensor structure of Higgs to
diboson couplings at the LHC (e.g., Ref.~\cite{Khachatryan:2014kca})
have largely used two formalisms.  One is
the specification of the most general $H \to VV$ amplitude allowed by the
relevant symmetries~\cite{Gao:2010qx,
    Bolognesi:2012mm}.
The second is to
specify the interaction Lagrangian terms consistent with these same
symmetries~\cite{Gainer:2013rxa, Gainer:2014hha}.  Two important points are in order: (i) these approaches
have different strengths and (ii) the approaches are nearly, though not
completely, equivalent in the physics they parameterize, given certain
reasonable physical assumptions.

Strengths of the amplitude approach include a greater ability to
parametrize absorptive (\textit{i.e.}, complex) contributions
to the amplitude arising from loop-induced contributions from light
particles running in the loop.\footnote{
    The magnitude of such contributions can be constrained using even
    conservative assumptions about the total Higgs width, as discussed,
    for example in Ref.~\cite{Chen:2013waa}.
}
The Lagrangian approach, on the other hand, is more useful when
utilizing standard computational tools, such as
FeynRules~\cite{Christensen:2008py, Alloul:2013bka},
CalcHEP~\cite{Pukhov:2004ca, Belyaev:2012qa},
CompHEP~\cite{Boos:1994xb, Pukhov:1999gg, Boos:2004kh}, and
MadGraph~\cite{Stelzer:1994ta, Maltoni:2002qb, Alwall:2007st,
    Alwall:2011uj}.
It may also make the connection with more comprehensive physical
frameworks more straightforward.

Recently, attention has been drawn to a third approach based on the use of
pseudo-observables (PO)~\cite{Gonzalez-Alonso:2014eva,David:2015waa,POYR4}.  This approach,
which was employed at the CERN Large Electron Positron (LEP) Collider~\cite{Bardin:1999gt,
    ALEPH:2005ab}, represents a generalization of the amplitude approach and thus it shares its strengths. 
    However, by demanding that the amplitude
considered describe a transition between on-shell states, the
PO method resolves several potential ambiguities and,
in particular, guarantees the gauge invariance of the quantities that are being
measured.
This property will become increasingly important as
measurements increase in sensitivity, probing contributions from NLO
effects in the SM~\cite{ Bordone:2015nqa, Gomez-Ambrosio:2015jea,
    Greljo:2015sla, Greljo:2017spw}, as well as potential new physics. 
The PO approach covers also BSM scenarios that affect the $H\to 4\ell$ channel without two intermediate $Z$ or $\gamma$ states. A simple example is a heavy $Z'$ boson in $H\to Z Z'\to 4\ell$, but more complicated cases with additional heavy exotic intermediate states (possibly coupled in a non-universal way to SM fermions) are also covered by this approach.

Thus PO provide a useful third approach to parameterizing
Higgs couplings.  However, as there are also advantages to the
amplitude and the Lagrangian frameworks (besides the important fact that
their use by the experiments is well-established~\cite{Sirunyan:2017tqd, Aaboud:2017vzb}), 
it is useful to describe how the PO approach compares to these other methods.
We do that in this paper.  In Section~\ref{sec:theory} we discuss the theoretical
relationship between various parameterizations.  In particular, we specify the
translation between the conventions used in each of these
approaches in detail, so that results obtained in the context of one framework
can
be easily understood in the context of the other frameworks.
In Section~\ref{sec:projections}, we analyze the future sensitivity to Higgs pseudo-observables within specific benchmark scenarios with only two free parameters. We discuss in some detail the role of the different experimental inputs, namely the total number of events observed, the ratios between different channels and the kinematic distribution of the events. It is hoped that these results will be informative in their own right, but will also provide a useful template for future experimental studies.

\section{Frameworks, Conventions, and Translation}
\label{sec:theory}

\subsection{The ``Amplitude'' Approach}
\label{subsec:amplitude}

In the ``amplitude'' approach ~\cite{
    Choi:2002jk,
    Gao:2010qx, DeRujula:2010ys,
    Bolognesi:2012mm}
that generally has been used in four-lepton
studies, the most general form of the $1 \to 2$ vertex describing the
decay of the putative Higgs boson to vector bosons is specified.  (Of
course, for the Standard Model Higgs the decay to $Z$ bosons is the
most important, though interesting phenomenology can result from
decays to $Z\gamma^\ast$ and
$\gamma^\ast\gamma^\ast$~\cite{Chen:2014gka}.)	The resulting
expression can be used to obtain helicity amplitudes
for the $H \to VV$ decay;\footnote{
    Here and below we suppress explicit indications that the vector bosons are,
    in general, off-shell.}
these can be combined with helicity amplitudes\footnote{
    We assume that $Z$ and $\gamma$ couple to leptons as in the Standard Model.
    We include also a possible contribution from a $Z'$ boson. 
    We allow its coupling to leptons to be an arbitrary sum of vector and axial vector contributions.}
for the decay of the vector bosons to leptons to obtain the
amplitude, and ultimately the differential cross section, for $g g \to
    H \to VV \to 4\ell$ (or processes where the Higgs is produced via
other mechanisms).  Following Ref.~\cite{Khachatryan:2014kca} (and references therein) we can write the general form of this amplitude as
\begin{eqnarray}
    \label{eq:amplitude}
   A(H \to V_1 V_2) = \frac{i}{v} \left\{
    \left[ a_{1}^{ V_1 V_2}
        + \frac{\kappa_1^{ V_1 V_2} q_1^2 + \kappa_2^{ V_1 V_2}
            q_2^{2}}{\left(\Lambda_{1}^{ V_1 V_2} \right)^{2}} \right]
    m_{V_1}^2 \epsilon_1^* \cdot \epsilon_2^* \right. \\ \nonumber
    \left.+\, a_{2}^{V_1V_2}  f_{\mu \nu}^{*(1)}f^{*(2),\mu\nu}
    +\, a_{3}^{V_1V_2}   f^{*(1)}_{\mu \nu} {\tilde f}^{*(2),\mu\nu} \right\},
\end{eqnarray}
where $\epsilon_i$ is the polarization vector and $q_i$ is the
momentum of the gauge boson labelled ``$V_i$'' ($i = 1,2$).  The
contribution to the amplitude corresponding to the field strength
tensor is $f^{(i)\mu\nu} = \epsilon^\mu_i q^\nu_i - \epsilon^\nu_i q^\mu_i$, and its dual is given by 
$\tilde{f}^{(i)\mu\nu} = \frac{1}{2}\epsilon_{\mu\nu\alpha\beta} f^{(i),\alpha\beta}$. 
Symmetry considerations and gauge invariance require $\kappa_1^{ZZ}=\kappa_2^{ZZ}$ and $a_1^{Z\gamma},a_1^{\gamma\gamma},\kappa_1^{\gamma\gamma},\kappa_2^{\gamma\gamma},\kappa_1^{Z\gamma}=0$.

We can proceed to find the helicity amplitudes that correspond to
Eq.~(\ref{eq:amplitude}) using explicit expressions for the polarization vectors $\epsilon_1$
and $\epsilon_2$.
In the case of the four-charged-lepton final state, the relevant $H \to V_1V_2$
sub-amplitudes are those where $V_1V_2$ refers to $ZZ$, $Z\gamma$,
$\gamma\gamma$, $ZZ^\prime$, etc.  It is important to note that terms
in the general
amplitude that vanish for on-shell vector bosons do not necessarily
vanish for off-shell vector bosons.  In particular, one can have a
$q_{\gamma}^2 \epsilon_{\gamma}^* \epsilon_{Z}^*$ contribution, which,
of course, vanishes when the photon is on-shell.

In general, the ``constants'' in Eq.~(\ref{eq:amplitude}) can be
taken to be Lorentz invariant
functions of $q_1$ and $q_2$.  (When the intermediate gauge
bosons are identical, the function must also be invariant under the
exchange of the labels $V_1$ and $V_2$.)   An arbitrary analytic Lorentz
invariant function of momenta may be expressed as
\begin{equation}
    f(q_1, q_2)  = f_0 + \frac{1}{\Lambda^2} (f_{21} q^2_1 +
    f_{22} q^2_2 + f_{23} (q_1 + q_2)^2) + \mathcal{O}(\Lambda^{-4}),
\end{equation}
where we assume that $f(q_1, q_2)$ can be expanded in a Taylor
series.  When $V_1$ and $V_2$ are produced in the decay of an on-shell
boson, the situation considered in this work, we can set $(q_1 +
    q_2)^2 = m_H^2$, where $m_H$ is the mass of the (Higgs) boson that
decays to $V_1$ and $V_2$.  Thus we see that Eq.~(\ref{eq:amplitude}) is
written to include the momentum dependence of the $a_{1}^{ V V}$
through $\mathcal{O}(\Lambda^{-2})$ terms, while only the leading
constant is included for $a_{2}^{ V V}$ and $a_{3}^{ V V}$.  This
apparent inconsistency is resolved when we realize that the amplitude
structures associated with $a_{2}^{ V V}$ and $a_{3}^{ V V}$ involve
two additional powers of momenta relative to the $a_{1}^{ V V}$
structure ($\epsilon_1^* \cdot \epsilon_2^*$).	Therefore, if we
take the $a_{i}^{ V V}$, the $\kappa_i$, etc. to be true
constants, the expression in Eq.~(\ref{eq:amplitude}) includes
the full momentum dependence of the amplitude (in the case where the
Higgs is on-shell) containing up to two powers of gauge boson momenta.
We will see similar truncations in the cases of the ``Lagrangian'' and the
``PO'' approaches discussed below.

\subsection{The ``Lagrangian'' Approach}
\label{subsec:lagrangian}

The $H \to VV$ interactions can also be described in the Lagrangian
formalism.  For $H \to 4 \ell$, the relevant interaction Lagrangian is
\begin{eqnarray}
    \label{eq:lagrangian}
    \mathcal{L} \supset
    -\kappa_{1,ZZ} \frac{M_Z^2}{v} H Z_\mu Z^\mu
    -\frac{\kappa_{2,ZZ}}{2 v} H Z_{\mu\nu} Z^{\mu\nu}
    -\frac{\kappa_{3,ZZ}}{2 v} H Z_{\mu\nu} \tilde{Z}^{\mu\nu} \\ \nonumber
    +\frac{\kappa_{4,ZZ} M_Z^2}{M_H^2 v} \Box H Z_\mu Z^\mu
    +\frac{2\kappa_{5,ZZ}}{v} H Z_\mu \square Z^\mu \\ \nonumber
    -\frac{\kappa_{2,Z\gamma}}{2 v} H Z_{\mu\nu} F^{\mu\nu}
    -\frac{\kappa_{3,Z\gamma}}{2 v} H Z_{\mu\nu} \tilde{F}^{\mu\nu}
    +\frac{\kappa_{5,Z\gamma} M_Z^2}{M_H^2 v} H Z_\mu \partial_\nu
    F^{\mu\nu} \\ \nonumber
    -\frac{\kappa_{2,\gamma\gamma}}{2 v} H F_{\mu\nu} F^{\mu\nu}
    -\frac{\kappa_{3,\gamma\gamma}}{2 v} H F_{\mu\nu} \tilde{F}^{\mu\nu}
    \\ \nonumber
    -\kappa_{1,ZZ^\prime} \frac{M_Z^2}{v} H Z_\mu Z^{\prime\mu},
\end{eqnarray}
where $F_{\mu\nu} = \partial_\mu A_\nu - \partial_\nu A_\mu$ is the
QED field strength tensor; $Z_{\mu\nu} = \partial_\mu Z_\nu
    - \partial_\nu Z_\mu$ is defined by analogy.  The dual field strength
tensors $\tilde{F}$ and $\tilde{Z}$ are defined by (\textit{e.g.})
$\tilde{F}_{\mu\nu} = \frac{1}{2} \epsilon_{\mu\nu\alpha
        \beta}F^{\alpha\beta}$, where $\epsilon$ is the totally
antisymmetric Levi-Civita tensor.
With the exception of interactions involving a $Z^\prime$, this
Lagrangian contains all operators up to dimension five (six, if one
assumes that the operators involve the vacuum expectation value of the
Higgs field).
We note that there are five $HZZ$ operators in
Eq.~(\ref{eq:lagrangian}), but fewer operators for $HZ\gamma$,
$H\gamma\gamma$, and $HZZ^\prime$.  In the case of $HZ\gamma$ and
$H\gamma\gamma$, this is due to the requirements provided by QED gauge
invariance.
For the $Z^\prime$ we have included only the lowest
dimensional operator, as amplitudes involving the heavy $Z^\prime$ are
suppressed by the large mass in the propagator.  Had we chosen to
consider a light $Z^\prime$, we would need to include the additional
$HZZ^\prime$ operators
\begin{eqnarray}
    \label{eq:HZZprime}
    \mathcal{L} \supset
    -\frac{\kappa_{2,ZZ'}}{2 v} H Z_{\mu\nu} Z^{\prime\,\mu\nu}
    -\frac{\kappa_{3,ZZ'}}{2 v} H Z_{\mu\nu} \tilde{Z}^{\prime\,\mu\nu} \\ \nonumber
    +\frac{\kappa_{4,ZZ'} M_Z^2}{M_H^2 v} \Box H Z_\mu Z^{\prime\,\mu}
    +\frac{2\kappa_{5,ZZ'}}{v} H Z_\mu \square Z^{\prime\,\mu}, \\ \nonumber
\end{eqnarray}
the $HZ^\prime\gamma$ operators
\begin{eqnarray}
    \label{eq:HZprimeGamma}
    \mathcal{L} \supset
    -\frac{\kappa_{2,Z'\gamma}}{2 v} H Z^\prime_{\mu\nu} F^{\mu\nu}
    -\frac{\kappa_{3,Z'\gamma}}{2 v} H Z^\prime_{\mu\nu} \tilde{F}^{\mu\nu}
    +\frac{\kappa_{5,Z'\gamma} M_Z^2}{M_H^2 v} H Z^\prime_\mu \partial_\nu
    F^{\mu\nu}
\end{eqnarray}
and the $HZ^\prime Z^\prime$ operators
\begin{eqnarray}
    \label{eq:HZprimeZprime}
    \mathcal{L} \supset
    -\kappa_{1,Z'Z'} \frac{M_Z^2}{v} H Z^\prime_\mu Z^{\prime\,\mu}
    -\frac{\kappa_{2,Z'Z'}}{2 v} H Z^\prime_{\mu\nu} Z^{\prime\,\mu\nu}
    -\frac{\kappa_{3,Z'Z'}}{2 v} H Z^\prime_{\mu\nu} \tilde{Z}^{\prime\,\mu\nu} \\
    \nonumber
    +\frac{\kappa_{4,Z'Z'} M_Z^2}{M_H^2 v} \Box H Z^\prime_\mu Z^{\prime\,\mu}
    +\frac{2\kappa_{5,Z'Z'}}{v} H Z^\prime_\mu \square Z^{\prime\,\mu}. \\ \nonumber
\end{eqnarray}
However, assuming the $Z^\prime$ mass to be comparable to the cutoff
of the theory, we can take all of the operators in
Eqs.~(\ref{eq:HZZprime}--\ref{eq:HZprimeZprime})
to be higher dimensional and therefore ignore them.\footnote{
    An alternate and perhaps clearer approach would be to
    integrate out the heavy $Z^\prime$.  We have chosen the approach above,
    with explicit $\kappa_{1,ZZ^\prime}$ term in the interest of consistency
    with parameterizations already in use by the experiments.}
For \textit{on-shell} Higgs studies (which are our focus here), the
$(\Box H) Z_\mu Z^\mu$ operator
can be absorbed into the $H Z_\mu
    Z^\mu$ operator, though this will not be the case when the Higgs is
off-shell~\cite{Gainer:2014hha}.
With the caveat that the coefficients
$\{\kappa_{i,VV}\}$ in Eq.~(\ref{eq:lagrangian}) are real, there is a
one-to-one mapping between the coefficients in Eqs.~(\ref{eq:lagrangian}) and
(\ref{eq:amplitude}), as shown explicitly in Table~\ref{tab:dictionary} (see also Ref.~\cite{Gainer:2014hha}).

While we assume that $Z$ bosons and photons couple to leptons
as in the SM, we can make no such assumptions about the interaction of the
$Z^\prime$ with leptons.
Thus, to fully characterize $H\to4\ell$ decays, we parameterize the
$Z^\prime$ interaction to SM leptons with
\be
\cL \supset  \sum_{\ell = \ell_L, \ell_R}  g_{Z^\prime}^\ell  \bar \ell
\gamma^\mu \ell Z^\prime_\mu~.
\ee

\subsection{The ``Pseudo-Observables'' Approach}
\label{subsec:pseudo-observables}

The PO formalism described here is based on the specification of the
pole structure of the amplitude for the entire $H \to 4\ell$ process. 
As we are considering an on-shell Higgs, and we single out the pole terms due to the
exchange of the SM gauge bosons, the resulting amplitude is automatically gauge invariant,
which is an advantage with respect to the ``amplitude'' approach described in
Subsection~\ref{subsec:amplitude}, especially when considering the
effects of higher order corrections~\cite{Bordone:2015nqa}.\footnote{
    If we were to consider contributions to high invariant mass
    four-lepton events from an off-shell Higgs boson~\cite{
        Caola:2013yja,
        Campbell:2013una,
        Gainer:2014hha, Kauer:2012hd,
        Kauer:2013cga, Campbell:2013wga,
        Passarino:2013bha, Englert:2014aca,
        Brivio:2014pfa, Cacciapaglia:2014rla,
        Azatov:2014jga, Campbell:2014gua,
        Englert:2014ffa,
        Buschmann:2014sia, Logan:2014ppa,
        Kauer:2015pma, Liebler:2015aka,
        Li:2015jva, Englert:2015bwa,
        Campanario:2015nha},
    we would
    need to construct PO for the whole $gg \to 4\ell$
    process, including both the ``signal'' $gg \to H \to VV \to 4\ell$
    amplitude and the ``background'' $gg \to VV \to 4\ell$ amplitudes with
    which it interferes.}
Furthermore, the PO formalism, like the amplitude
formalism, can accommodate arbitrary loop corrections in a relatively
straightforward way.

Following Ref.~\cite{Gonzalez-Alonso:2014eva}, the goal of $H\to 4\ell$ studies
in the PO approach is to characterize in general terms the correlation function
of the Higgs and two fermion currents,
\be
\langle 0 | {\cal T} \{ H(0), J_\ell^\mu(x), J_{\ell'}^\nu(y) \} | 0
\rangle~,
\label{eq:CorrFunc}
\ee
where $\ell, \ell' = e,\mu$.
Assuming Lorentz invariance and reasonable assumptions about flavor
conservation, the matrix element corresponding to this correlation function can
be described by
\begin{eqnarray}
    && \mathcal{A}_{n.c.} \left[ X \to  \ell^- (p_1) \, \ell^+ (p_2) ,
        \ell^{\prime -} (p_3) \, \ell^{\prime +} (p_4) \right]	=
    i \frac{2 m^2_Z}{ v} \sum_{\ell = \ell_L, \ell_R}
    \sum_{\ell' = \ell'_L, \ell'_R} (\bar \ell  \gamma_\rho \ell)
    (\bar \ell'  \gamma_\sigma \ell')
    \mathcal{T}^{\rho\sigma} (q_1 ,q_2)  \nonumber \\
    &&
    \mathcal{T}^{\rho\sigma} (q_1, q_2) =  \bigg[
    F^{\ell\ell'}_1 (q_1^2, q_2^2) g^{\rho\sigma} +
    F^{\ell\ell'}_3 (q_1^2, q_2^2)	\frac{ {q_1} \cdot
    {q_2}~g^{\rho\sigma} - {q^\rho_2} {q^\sigma_1} }{m_Z^2}
    \nonumber
    \\
    &&~~~~~~~~~~~~~~~~~~~~
    +F^{\ell\ell'}_4 (q_1^2, q_2^2)
    \frac{\varepsilon^{\rho\sigma\alpha\beta} q_{V_2\,\alpha} q_{V_1\,\beta}
    }{m_Z^2} \bigg],
    \label{eq:h4l1}
\end{eqnarray}
where $q_1=p_1 +p_2$ and $q_2=p_3 +p_4$.
Recognizing the presence of physical poles in the correlation function
\eqref{eq:CorrFunc} due to the
propagation of intermediate SM gauge bosons, we expand around these poles and
define the PO
directly from their residues:
\begin{eqnarray}
    \label{F1}
    F^{\ell\ell'}_1 (q_1^2, q_2^2) &=&
    \kappa_{ZZ}  \frac{ g_Z^\ell g_Z^{\ell'}  }{P_Z(q_1^2) P_Z(q_2^2)} +
    \frac{\epsilon_{Z \ell}}{m_Z^2}  \frac{ g_Z^{\ell'} }{P_Z(q_2^2)} +
    \frac{\epsilon_{Z \ell'}}{m_Z^2} \frac{g_Z^\ell}{  P_Z(q_1^2)}
    \\ && +
    \Delta^{\rm SM}_{1} (q_1^2, q_2^2)
    ~,   \label{eq:h4lNeutrCurrEFT1}  \nonumber \\
    \label{F3}
    F^{\ell\ell'}_3 (q_1^2, q_2^2) &=&
    \epsilon_{ZZ}  \frac{ g_Z^\ell g_Z^{\ell'}  }{P_Z(q_1^2) P_Z(q_2^2)} -
    \epsilon_{Z\gamma}
    \left(
    \frac{  e  g_Z^\ell   }{ q_2^2  P_Z( q_1^2) }   +
    \frac{ e g_Z^{\ell'}   }{ q_1^2  P_Z( q_2^2) }
    \right)
    \label{eq:h4l2b} \\ &&
    +\, \frac{\epsilon_{\gamma\gamma}	e^2}{ q_1^2 q_2^2  }
    \nonumber +\Delta^{\rm SM}_{3} (q_1^2, q_2^2),
    \\ \label{F4}
    F^{\ell\ell'}_4 (q_1^2, q_2^2) &=&
    \epsilon^{\rm CP}_{ZZ}	\frac{
    g_Z^\ell  g_Z^{\ell'} }{P_Z(q_1^2) P_Z(q_2^2)} -
    \epsilon^{\rm CP}_{Z\gamma}  \left(
    \frac{  e g_Z^\ell   }{ q_2^2  P_Z( q_1^2) }	+
    \frac{ e g_Z^{\ell'}   }{ q_1^2  P_Z( q_2^2) }	\right)
    \\ && \nonumber
    +\, \frac{\epsilon^{\rm CP}_{\gamma\gamma} e^2 }{ q_1^2 q_2^2  }~,~
    \label{eq:h4l2}
\end{eqnarray}
where $g_Z^{\ell,\ell'}$ correspond to the well-measured LEP-I $Z$-pole PO,
\be
\mathcal{A}(Z(\varepsilon) \to \ell^+  \ell^-)
=  i \sum_{\ell=\ell_L,\, \ell_R}   ~ g_Z^{\ell} \, \varepsilon_\nu
\, \bar \ell \, \gamma^\nu \ell,
\label{eq:gZf}
\ee
and  $P_Z(q^2) =  q^2 -m_Z^2 + i m_Z \Gamma_Z$.
Note the minus signs in Eqs.~(\ref{F3}) and (\ref{F4}) occur because the
relevant expressions contain a $Q$ for the fermion charge in
Ref.~\cite{Gonzalez-Alonso:2014eva}, which is equal to minus one for electrons and muons. 
In this decomposition around physical poles, we neglected further terms which 
are necessarily generated by local operators of dimension greater
than 6 and thus strongly suppressed if the new physics scale is above the
electroweak scale.
The parameters $g_Z^{f}$, $\kappa_{ZZ}$, and $\epsilon_X$
are thus well-defined PO; they can be measured in experiments and their values
can be calculated at any given order in perturbation theory in any
underlying theory (or effective field theory).\footnote{
    Here we generically denote by $\epsilon_X$ the parameters
    $\epsilon_{ZZ,Z\gamma, \gamma\gamma, Zf}$ and 
    $\epsilon^{\rm CP}_{ZZ, Z\gamma, \gamma\gamma}$.}

While the amplitude for $H \to 2e2\mu$ is given directly by Eq.~\eqref{eq:h4l1},
the decays $H \to 4e, 4\mu$ include an interference between two contributions,
corresponding to the two possible assignments of the fermions inside the
currents:
\ba
\cA \left[ h \to  \ell (p_1) \bar \ell (p_2) \ell  (p_3) \bar \ell (p_4)
    \right] &=&
\cA_{n.c.} \left[ h \to  f (p_1) \bar f (p_2) f^\prime	(p_3) \bar f^\prime
    (p_4) \right]_{f=f^\prime=\ell} \no \\
& - &	\cA_{n.c.} \left[ h \to  f (p_1) \bar f (p_4) f^\prime	(p_3) \bar
    f^\prime (p_2) \right]_{f=f^\prime=\ell}~.
\ea
The tree-level connection between the PO and the parameters used in
Eqs.~(\ref{eq:amplitude}) and (\ref{eq:lagrangian}) is given in
Table~\ref{tab:dictionary}.  Of these parameters, only $\epsilon_{Z e_{L,R}}$
and $\epsilon_{Z \mu_{L,R}}$ can depend on the flavor of the final-state
leptons.  In the limit, where we neglect loop contributions from light
states, $\kappa_{ZZ}$ and $\epsilon_X$ are all real.
The functions  $\Delta^{\rm SM}_{1,3} (q_1^2, q_2^2)$ describe
loop-induced SM contributions, which cannot be described in terms of $D \leq 6$ 
effective operators; an explicit expression for 
$\Delta_3^{\rm SM}(q_1^2, q_2^2)$ at one-loop
can be found in Ref.~\cite{Gonzalez-Alonso:2014eva}, where it is shown
that their size is small and thus can be neglected given experimental precision
in the foreseeable future.

\begin{table}[t]
\centering
\begin{tabular}{| c | c | c |}
\hline 
PO & Lagrangian parameter & Amplitude parameters\\ \hline
$\kappa_{ZZ}$ & $-\kappa_{{1,ZZ}}-\kappa_{4,ZZ}-2\kappa_{5,ZZ}$ &  $\frac{1}{2}a_1^{ZZ} + \frac{m_Z^2}{(\Lambda_1^{ZZ})^2}\kappa_{1}^{ZZ} $ \\
$\epsilon_{ZZ}$ & $\kappa_{2,ZZ}$ & $ a_2^{ZZ} $ \\
$\epsilon^{\rm CP}_{ZZ}$ & $\kappa_{3,ZZ}$  & $ a_3^{ZZ}$  \\ 
$\epsilon_{Z\gamma}$   &  $\kappa_{2,Z\gamma}/2$ & $a_2^{Z\gamma}$ \\
$\epsilon^{\rm CP}_{Z\gamma}$ &  $\kappa_{3,Z\gamma}/2$ & $a_3^{Z\gamma}$ \\
$\epsilon_{\gamma\gamma}$ &  $\kappa_{2,\gamma\gamma}$ & $a_2^{\gamma\gamma}$ \\
$\epsilon^{\rm CP}_{\gamma\gamma}$ & $\kappa_{3,\gamma\gamma}$ & $a_3^{\gamma\gamma}$ \\ 
$\epsilon_{Z f} $ & $ -g_{Z}^f \, \kappa_{{5,ZZ}}+g_{Z^\prime}^f \, \frac{\kappa_{{1,ZZ^\prime}}}{2}\frac{m_Z^2}{m_{Z^\prime}^2}-e\,\frac{\,\kappa_{{5,Z\gamma}}}{2} \frac{m_Z^2}{m_H^2}$ & $ g_{Z}^f \frac{\kappa_{1}^{{ZZ}}~m_Z^2}{ 2(\Lambda_1^{{ZZ}})^2}~- g_{Z^\prime}^f \frac{a_{1}^{{ZZ'}}~m_Z^2}{2\,m_{Z^\prime}^2}- e \frac{\kappa_{2}^{{Z\gamma}}~m_Z^2}{2(\Lambda_1^{{Z\gamma}})^2}$ \\
\hline \hline
\end{tabular}
\caption{\label{tab:dictionary}
In this table we provide a ``dictionary'', allowing one to convert (at the tree level)
between (i) the ``amplitude'' formalism of Eq.~(\ref{eq:amplitude}),
(ii) the ``Lagrangian'' formalism of Eq.~(\ref{eq:lagrangian}), and
(iii) the ``pseudo-observables'' formalism of Eq.~(\ref{eq:h4l1}).
The label $f$ is a generic label for a fermion ($f=e_L$, $e_R$, $\mu_L$, $\mu_R$).}
\end{table}

The tree-level matching of the operators in the Lagrangian approach to the
PO, which is shown in Table~\ref{tab:dictionary}, is obtained by computing the same decay amplitude and rearranging the
terms so that the expansion around physical poles is recovered.
For example, the $H Z_\mu \partial_\nu F^{\mu\nu}$ operator in
Eq.~(\ref{eq:lagrangian}) corresponds to an amplitude with a
$q^2_{\gamma}$ in the numerator (from the $\Box$ operator).  This
$q^2_{\gamma}$ cancels the photon propagator in the denominator, so
this operator contributes to $\epsilon_{Z e}$ and $\epsilon_{Z \mu}$.
The contribution of the heavy $Z'$ is evaluated in the limit $q^2_V \ll m_{Z^\prime}^2$ (consistently with our general assumptions).
In such limit the $Z'$ can easily be integrated out, generating a  contribution to the contact terms with the flavor structure given by its couplings to fermions.
Let us stress, however, that the $Z'$ is just a specific example of a BSM scenario that generates flavor-dependent contact terms. The PO approach is not restricted in any sense to this particular case, in contrast to the Amplitude or Lagrangian approaches.


\subsection{Global Approaches to Higgs Couplings}

While our interest in this paper is in detailed parameterizations of BSM effects in $H\to 4\ell$, we note in passing that much work
has gone into ``global'' studies of Higgs couplings to
various final states, whether in the so-called ``$\kappa$'' or
``signal-strength'' formalism~\cite{Azatov:2012bz, Klute:2012pu,
    Montull:2012ik, Espinosa:2012im, Carmi:2012in, Freitas:2012kw,
    Giardino:2013bma,
    Ellis:2013lra}) or in the context of SM effective
theories~\cite{Englert:2014aca, Buchmuller:1985jz,
    Hagiwara:1986vm,
    Giudice:2007fh, Grzadkowski:2010es, Bonnet:2011yx, Corbett:2012dm,
    Ellis:2012hz, Cacciapaglia:2012wb, Masso:2012eq, Belanger:2012gc,
    Corbett:2012ja, Grojean:2013kd, Cheung:2013kla, Elias-Miro:2013gya,
    Falkowski:2013dza, Contino:2013kra,
    Djouadi:2013qya, Corbett:2013pja, Dumont:2013wma, Bechtle:2013xfa,
    Belanger:2013xza, Chpoi:2013wga, Elias-Miro:2013mua,
    Pomarol:2013zra, Boos:2013mqa, Alloul:2013naa, Delgado:2013hxa,
    Willenbrock:2014bja, Englert:2014uua, Ellis:2014dva,
    Masso:2014xra, Biekoetter:2014jwa, deBlas:2014ula, Englert:2014ffa,
    Cheung:2014oaa,
    Goertz:2014qta, Ellis:2014jta,
    Bergstrom:2014vla, Corbett:2014ora, Gonzalez-Fraile:2014cya,
    Cheung:2015uia, Gorbahn:2015gxa, Falkowski:2015fla, Falkowski:2015jaa, Ghezzi:2015vva,
    Corbett:2015ksa, Dwivedi:2015nta, Fichet:2015xla, Gregersen:2015uea,
    Englert:2015hrx, Drozd:2015rsp, Reina:2015yuh, Butter:2016cvz,
    Brivio:2016fzo, Dawson:2016ugw,
    Bauer:2016hcu, Freitas:2016iwx, Englert:2016ljt, Ciuchini:2016sjh,
    Corbett:2017ieo,
    Corbett:2017qgl, Brivio:2017vri, Helset:2017mlf,
    Dedes:2018seb}.
While such studies are clearly of
value, we wish to emphasize that our goal here is quite different.
Specifically, our goal is to allow a description of BSM effects in $H\to 4\ell$ that is as theoretically unbiased as possible using as much information
as possible.

Generally, more global approaches to Higgs couplings
make numerous additional assumptions and use less information about a
specific process, such as Higgs to four leptons, than is
experimentally available.  As the assumptions in these studies tend to
be theoretically reasonable, such work is extremely useful in
providing a ``big picture'' view of what many different channels are
telling us about the Higgs, which is complementary to our goal of
extracting the most possible information from a single channel.  
To give an indication of what is possible in such analyses and to
illustrate the use of the various formalisms described here,
particularly the PO framework presented in
Subsection~\ref{subsec:pseudo-observables}, we present projections, in
Section~\ref{sec:projections},
for the sensitivity of LHC measurements in these frameworks.

\section{LHC Projections}
\label{sec:projections}

\subsection{Benchmark scenarios}
\label{subsec:scenarios}

While the PO are directly related to physical properties of the
Higgs decay
amplitudes, the generic expectation is that explicit new physics models would 
contribute to some combination of PO. For this reason, the experimental analysis
should be performed with the most general case possible in mind. In order to 
reduce the number of independent parameters, the best option is to impose 
relations due to specific symmetries \cite{Gonzalez-Alonso:2014eva}.

With this important caveat in mind, but also with an eye to studying the
sensitivity of the LHC and other future colliders for measuring Higgs PO in 
$H \to 4\ell$ decays, let us list the following six simplified scenarios, each of 
which has, conveniently, only two independent PO:
\begin{enumerate}
    \item {\bf  Longitudinal vs. transverse:} ($\kappa_{ZZ}$ vs.
          $\epsilon_{ZZ}$), where all other $\epsilon \to 0$.
          \label{scen1}
    \item {\bf  CP admixture:} ($\kappa_{ZZ}$ vs. $\epsilon^{\rm CP}_{ZZ}$),
          where all other $\epsilon \to 0$.
          \label{scen2}
    \item {\bf  Linear EFT-inspired \cite{Gonzalez-Alonso:2015bha}:}
          ($\kappa_{ZZ}$ vs. $\epsilon_{Z \ell_R}$), where
          $\epsilon_{Z \ell_L} = 2 \epsilon_{Z \ell_R}$,
          $\epsilon_{Z e_{L,R}} = \epsilon_{Z \mu_{L,R}}$, and other
          $\epsilon \to 0$ (or $\kappa_{ZZ}$ vs. some other combination of
          contact terms).
          \label{scen3}
    \item {\bf  Flavor universal contact terms:} ($\epsilon_{Z \ell_L}$ vs.
          $\epsilon_{Z \ell_R}$), where $\epsilon_{Z e_{L,R}} = \epsilon_{Z \mu_{L,R}}$,
          $\kappa_{ZZ} = 1$, and other $\epsilon \to 0$.
          \label{scen4}
    \item {\bf  Flavor non-universal vector contact terms:} ($\epsilon_{Z e_R}$
          vs. $\epsilon_{Z \mu_R}$), where $\epsilon_{Z e_L} = \epsilon_{Z e_R}$,
          $\epsilon_{Z \mu_L} = \epsilon_{Z \mu_R}$, $\kappa_{ZZ} = 1$, and other
          $\epsilon \to 0$.
          \label{scen5}
    \item {\bf  Flavor non-universal axial contact terms:} ($\epsilon_{Z e_R}$
          vs. $\epsilon_{Z \mu_R}$), where $\epsilon_{\ell_L} = - \epsilon_{Z \ell_R}$,
          $\kappa_{ZZ} = 1$, and other $\epsilon \to 0$.
          \label{scen6}
\end{enumerate}
Scenarios 1 and 2 have been the focus of much of the existing work in four-lepton 
phenomenology and experiment, whereas the rest of cases have received much less attention. 
Scenarios 1, 2, and 3 include $\kappa_{ZZ}$, which only affects the overall normalization of all $H\to4\ell$ channels and which consequently can only be probed through its effect in the total rates. This in turn requires the implicit assumption that other BSM effects affecting Higgs production are absent. For this reason in the next subsections we focus on the last three scenarios, where kinematic distributions and ratios of rates also provide interesting information on all PO under study. This study will serve as an example of the use of the PO formalism. 
These scenarios all involve contact operators that couple the Higgs to an
intermediate boson and two leptons. They could arise, {\it e.g.}, via a heavy $Z^\prime$ 
that is integrated out, but also via other mechanisms that might not even involve two intermediate bosons. 
In scenario~\ref{scen4}, this interaction is of the same 
strength for electrons and muons; the helicity structure of the coupling gives 
us two independent parameters.  In scenarios~\ref{scen5} and \ref{scen6}, we fix
the helicity 
structure of the couplings to be vector or axial, respectively, and
make the couplings to electrons and muons the two independent parameters. The violation of lepton-flavor universality makes these two scenarios qualitatively different from the rest, both conceptually and in practice, as we will see in the phenomenological analyses below.

\subsection{Projections Using Rates and Kinematics}
\label{subsec:ATLASapproach}

The ATLAS collaboration has recently published the first $h \to 4 \ell$ analysis in the PO framework using 36.1~fb$^{-1}$ of data at 13~TeV~\cite{Aaboud:2017oem} (see also~\cite{Mancini:2016flk}).
They work in the (lepton flavor universal) benchmark scenarios 3 and 4, using the binned $(m_{12},m_{34})$ invariant mass distribution as experimental input. Let us stress that they do not work with normalized distributions, and thus they are not only sensitive to the effects of the PO in the shapes but also in the total rates. As mentioned above, this assumes SM-like Higgs production, which allows them to probe $\kappa_{ZZ}$ in scenario 3. 

In this section we expand the scope of this ATLAS search and go one step beyond, highlighting the importance of lepton flavor universality tests. Namely, we split the $(m_{12},m_{34})$ histogram, shown in Fig.~3 (right) of~\cite{Aaboud:2017oem}, into four categories based on the lepton flavor: $4e$, $4\mu$, $2e2\mu$, and $2\mu2e$.\footnote{The channels $2e2\mu$ and $2\mu2e$ are split according to which lepton pair comes from the onshell Z boson, or more precisely, which one has an invariant mass closest to the $Z$ mass.} Apart from that, we closely follow the ATLAS analysis of Ref.~\cite{Aaboud:2017oem} and obtain projections for our scenarios 4, 5 and 6.

Let us first explain our simulation procedure. The signal events ($p p \to h \to 4 \ell$) are generated using the {\tt HiggsPO} UFO model from Ref.~\cite{Greljo:2017spw} within the {\tt MadGraph5\_aMC@NLO} framework~\cite{Alwall:2014hca}. Reliable gluon-gluon fusion production kinematics is obtained with the leading order 
matrix element and parton shower jet merging, while the normalization factor ($K_F = 2.32$) is taken from the best higher-order QCD prediction~\cite{Anastasiou:2016cez}. Subsequent showering and hadronisation effects are simulated with {\tt Pythia 6}~\cite{Sjostrand:2014zea}, while the detector effects are simulated with {\tt Delphes 3}~\cite{Ovyn:2009tx}. Event samples are generated for enough points in the PO parameter space allowing for the reconstruction of the quadratic dependence of any observable (to be discussed below). The dominant background, coming from $p p \to Z Z^*(\gamma^*) \to 4 \ell$, follows a similar simulation pipeline. We estimate the NLO QCD effects for the background by computing the $K$-factor in the signal region ($K_F = 1.3$), and using it to rescale the LO simulation. 

We closely follow the event selection of the ATLAS search. The signal region is defined as follows. Four leptons ($\ell = e, \mu$) are selected to make a lepton quadruplet. Electrons are required to have $E_T > 7$~GeV and $\eta<2.47$, while muons satisfy $p_T > 5$~GeV and $\eta<2.7$. Jets are reconstructed with the anti-$k_T$ algorithm and considered when $p_T > 30$~GeV and $\eta<4.5$. A lepton quadruplet consisting of two pairs of same flavor opposite-charge leptons is required, with the $p_T$ cuts in the quadruplet of 20~GeV, 15~GeV and 10~GeV for the leading leptons. Quadruplets with same flavor opposite-charge lepton's invariant mass below $5$ GeV are discarded. The opposite sign same flavor lepton pair closest to the $Z$-boson mass is referred as the leading dilepton, with invariant mass, $m_{12}$, required to be between 50~GeV and 106~GeV. The sub-leading dilepton invariant mass, $m_{34}$, is required to be in the range 12~GeV to 115~GeV. As a validation of our recast procedure, we correctly reproduce the ATLAS expected signal and background events in Fig.~3 (right) of Ref.~\cite{Aaboud:2017oem}.

\begin{figure}[t]
    \begin{center}
      \includegraphics[width=0.325\textwidth]{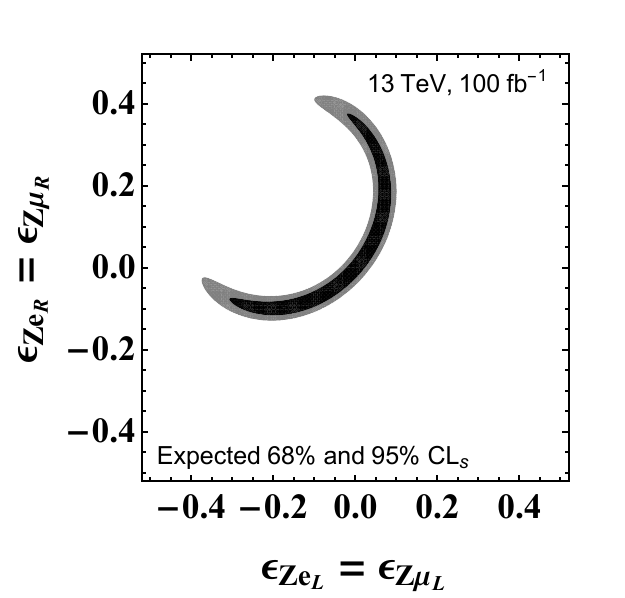}
      \includegraphics[width=0.325\textwidth]{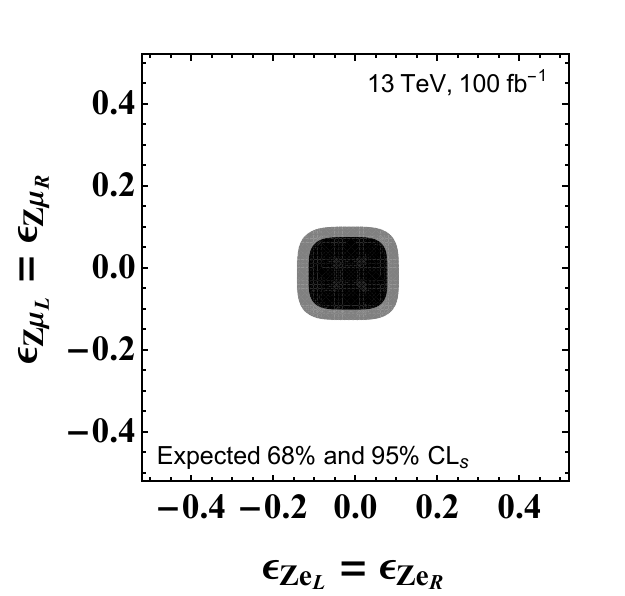}
      \includegraphics[width=0.325\textwidth]{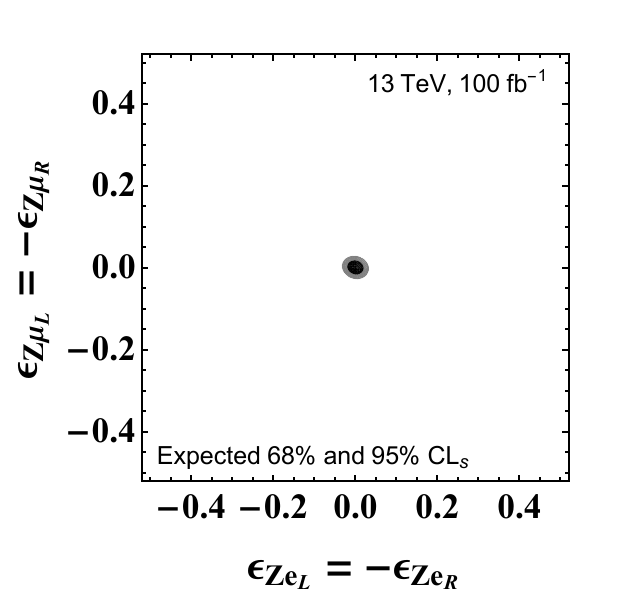}
    \end{center}
    \vspace{-10 pt}
    \caption{
       Projection of the ATLAS analysis~\cite{Aaboud:2017oem} to $100$~fb$^{-1}$ ($\sqrt{s}=13$ TeV) for scenarios 4, 5, and 6. Black (gray) regions represent $68\%$ ($95\%$) expected CL intervals.
        \label{fig:asn}}
\end{figure}

The number of events obtained in the each bin is proportional to the corresponding squared matrix element, which is a quadratic function of the PO. 
Therefore, we introduce matrices, $X_{ij}$, for every bin to express the number of events as
\begin{equation}
\frac{N}{N^{SM}}=\sum_{j \ge i} X_{ij} \epsilon_i  \epsilon_j\,,
\label{eq:NiX}
\end{equation}
where $N^{SM}$ is the number of expected events in the SM and where we introduced the vector $\epsilon \equiv \left(\epsilon_{Z e_{L}},\epsilon_{Z e_{R}},\epsilon_{Z \mu_{L}},\epsilon_{Z \mu_{R}}\right)^T$. The latter only contains contact terms because we focus on the benchmark scenarios 4, 5 and 6. The extension to a more general case with more PO is straightforward. In order to determine the $X$ matrices, we need to run the simulation for different values of the Higgs PO as the input parameters. This needs to be done ten times at least, since this is the number of independent parameters of a real, symmetric, $4\times 4$ matrix. Furthermore, we divided the points in the $m_{12}-m_{34}$ plane into five bins, as was done in the ATLAS search~\cite{Aaboud:2017oem}. After we calculate the $X$ matrices, we can compare ``exotic PO" with the ``SM Higgs" by constructing the appropriate likelihood function. Although we will work with a fix luminosity of 100 fb$^{-1}$, let us note that the $X_{ij}$ coefficients do not scale with the luminosity.

The events are categorised by four decay channels, $h\to 4e$, $h\to 4\mu$, $h\to 2e2\mu$ and $h\to 2\mu 2e$, each further separated into five bins. The likelihood function is given by the product of Poisson probabilities for all bins and all categories:
\begin{equation}
L(\epsilon)=\prod_n\frac{\text{exp}(-\mu_n)(\mu_n)^{N_n^{\text{exp}}}}{N_n^{\text{exp}}!}, \label{eq:likelihood}
\end{equation}
where the index $n$ refers to a specific bin and category, and
\begin{equation}
\mu_n=N_n^{\text{bkg,SM}}+N_n^{\text{sig,SM}}~\epsilon^TX_n\epsilon.
\end{equation}
The quantities $N_n^{\text{bkg,SM}}$ and $N_n^{\text{sig,SM}}$ represent the number of events per category and bin for background and signal events within the SM, respectively. $N_n^{\text{exp}}$ is number of events per bin and category obtained in the experiment. For the purpose of this study, we neglect systematic uncertainties.

Using the likelihood function in Eq.~\eqref{eq:likelihood}, we construct a profile likelihood ratio statistic, $\lambda (\epsilon) \equiv - 2 \log (L(\epsilon) /L(\hat \epsilon))$, to obtain $68\%$ and $95\%$ expected confidence levels for $100$~fb$^{-1}$ assuming the SM. Our results for scenarios 4, 5 and 6, are shown in Fig.~\ref{fig:asn} from left to right, respectively. In the following, we will try to break down these limits and understand the impact of the overall normalization, lepton universality rate ratios and event kinematics separately.

\subsection{Projections Using Relative Normalization and Kinematics}
\label{subsec:CMSapproach}

In this section, we study again the sensitivity of future LHC data to Higgs pseudo-observables using benchmark scenarios~\ref{scen4}--\ref{scen6}. However, in contrast to the analysis in the previous subsection, we will not include the information contained in the overall normalization common to all $H\to4\ell$ flavor channels. Moreover, we will not use the binned $(m_{12},m_{34})$ invariant mass distribution but will instead use the unbinned one with full event-by-event information. This approach is similar to that followed in, {\it e.g.}, the CMS search described in Ref.~\cite{Khachatryan:2014kca}.  

In addition to presenting results derived using only kinematic information, we will also present results where the discovery significance shown results from both kinematical information and  ``relative normalization'', \textit{i.e.}, the relative number of $4e$, $4\mu$, and $2e2\mu$ events.
These analyses help us to understand how much of the sensitivity to exotic PO points is due to 
kinematical information, how much is due to changes in the ratio of various event types, and how much is due to 
the overall normalization.

\begin{figure}
    \centering
    \begin{subfigure}[b]{0.48\textwidth}
        \includegraphics[width=\textwidth]{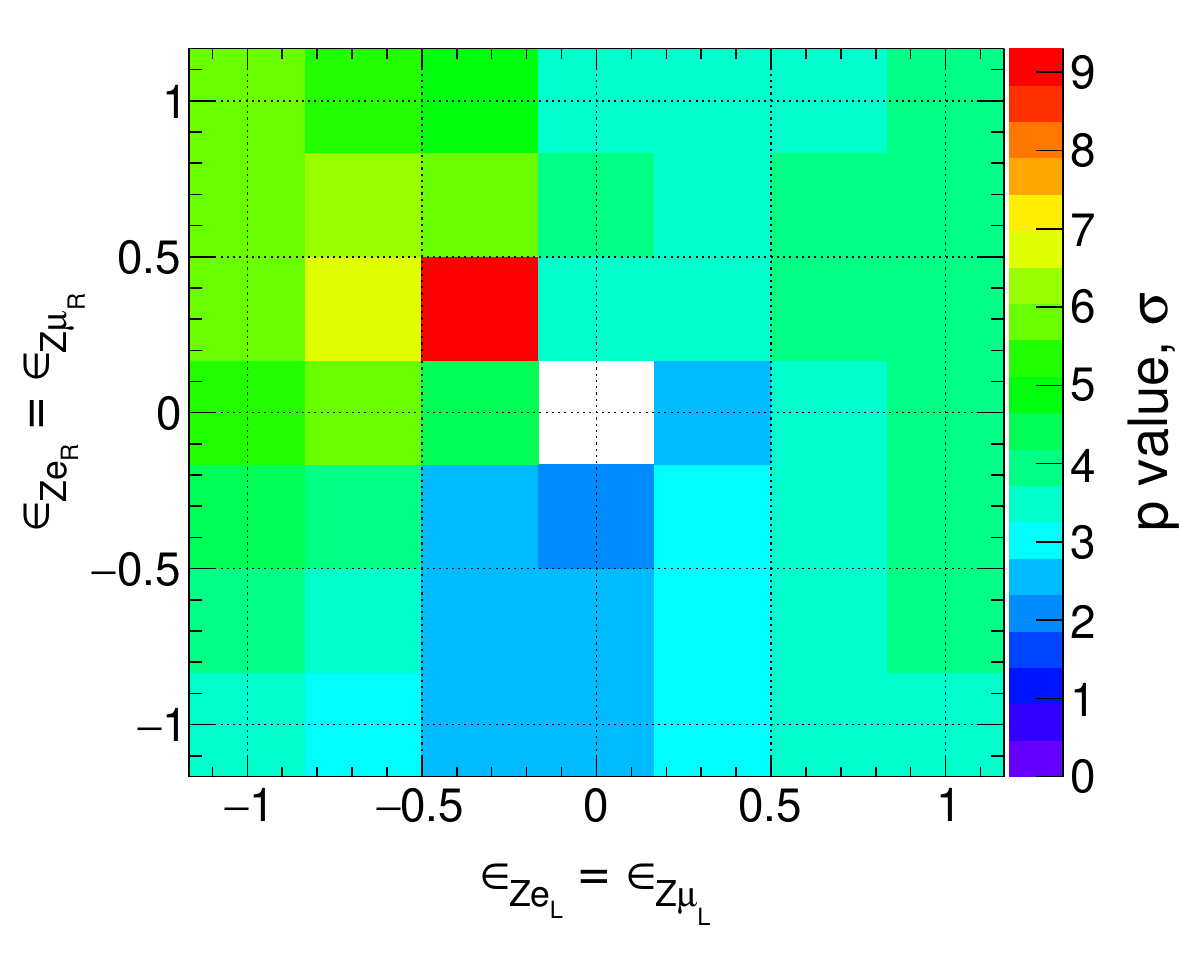}
        \caption{A full-range version}
        \label{fig:scen4a}
    \end{subfigure}
    ~ 
    \begin{subfigure}[b]{0.48\textwidth}
        \includegraphics[width=\textwidth]{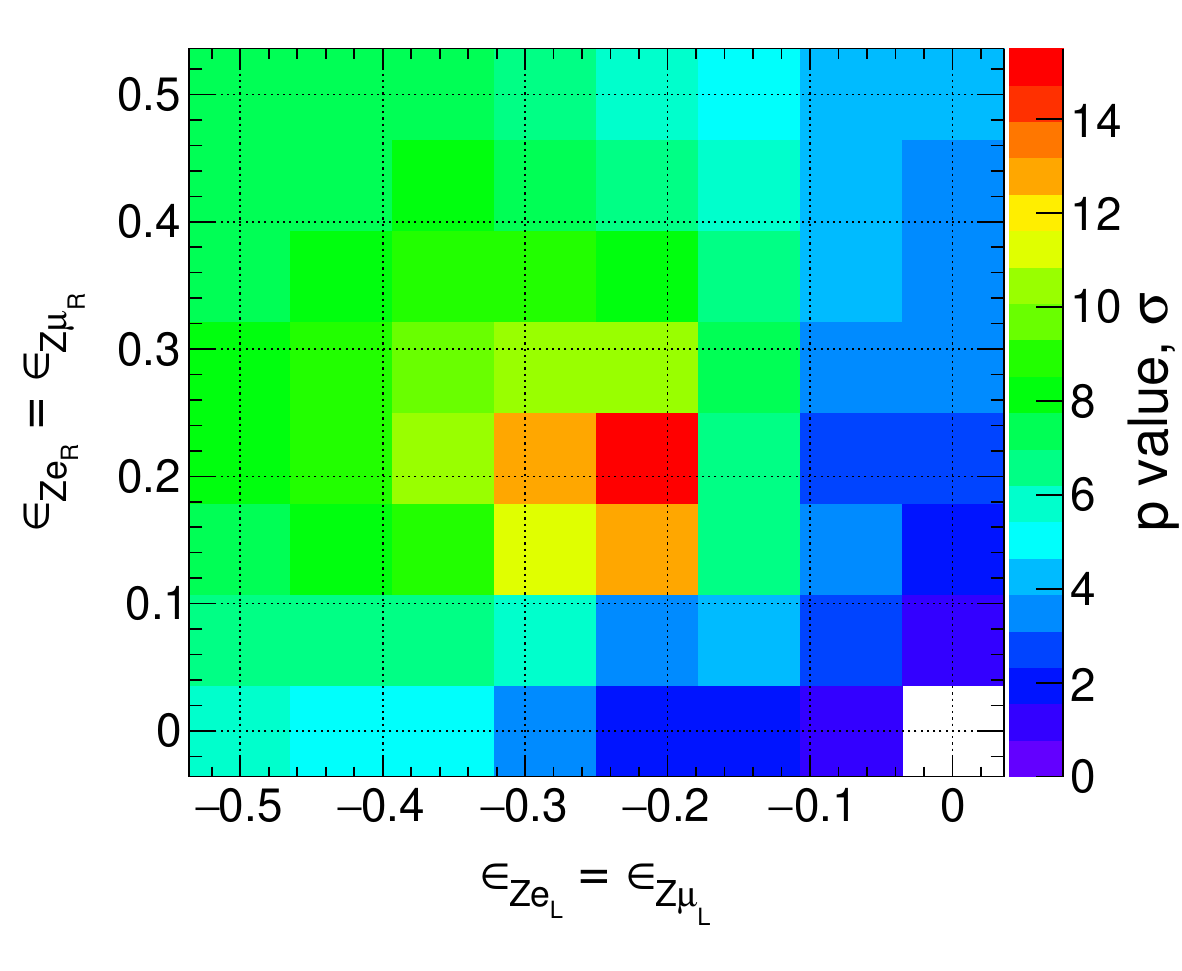}
        \caption{A zoomed-in version}
        \label{fig:scen4b}
    \end{subfigure}
    \caption{These figures show the expected discovery sensitivity in sigma (with 100 fb$^{-1}$ at 13 TeV) for
    the hypothesis described in scenario~\ref{scen4}, where, in addition to the
    SM coupling, the PO
    $\epsilon_{Z \ell_L}  = \epsilon_{Z e_L} = \epsilon_{Z \mu_L} $ and
    $\epsilon_{Z \ell_R}  = \epsilon_{Z e_R} = \epsilon_{Z \mu_R} $
    take on non-zero values.  These quantities are specified on the x and y axes;
    the SM is reflected by the point $(0, 0)$.
    The figure on the right, Fig.~\ref{fig:scen4b}, gives a zoomed-in version of
    the second quadrant of Fig.~\ref{fig:scen4a}.
    The results are obtained using only the kinematical distribution of the events via the procedure specified in 
    Subsection~\ref{subsec:CMSapproach}. Including  information about the relative normalization of the various $H\to4\ell$ channels has a relatively small impact here.}
    \label{fig:scen4}
\end{figure}

\begin{figure}[t]
    \centering
    \begin{subfigure}[b]{0.48\textwidth}
        \includegraphics[width=\textwidth]{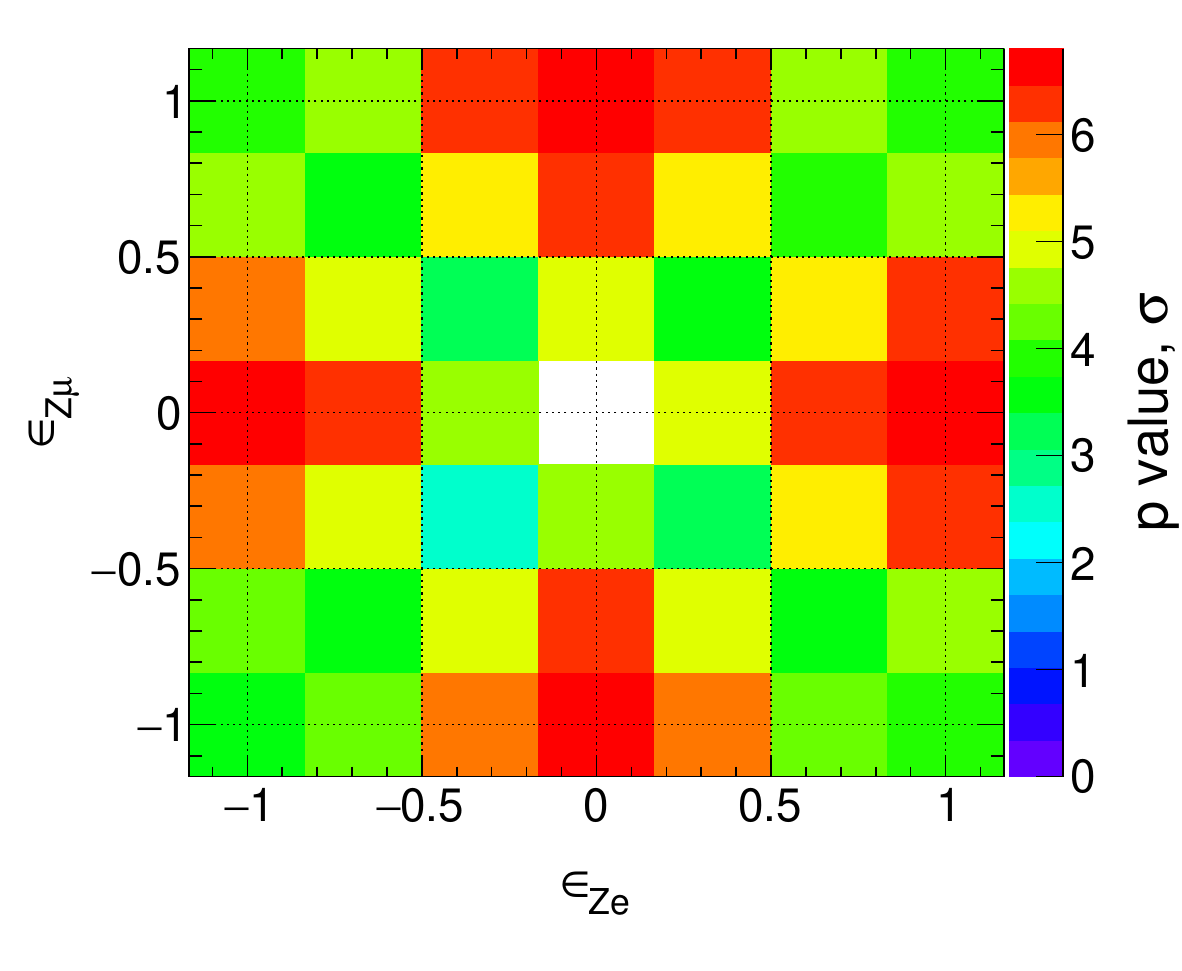}
        \caption{Kinematics}
        \label{fig:scen5a}
    \end{subfigure}
    ~ 
    \begin{subfigure}[b]{0.48\textwidth}
        \includegraphics[width=\textwidth]{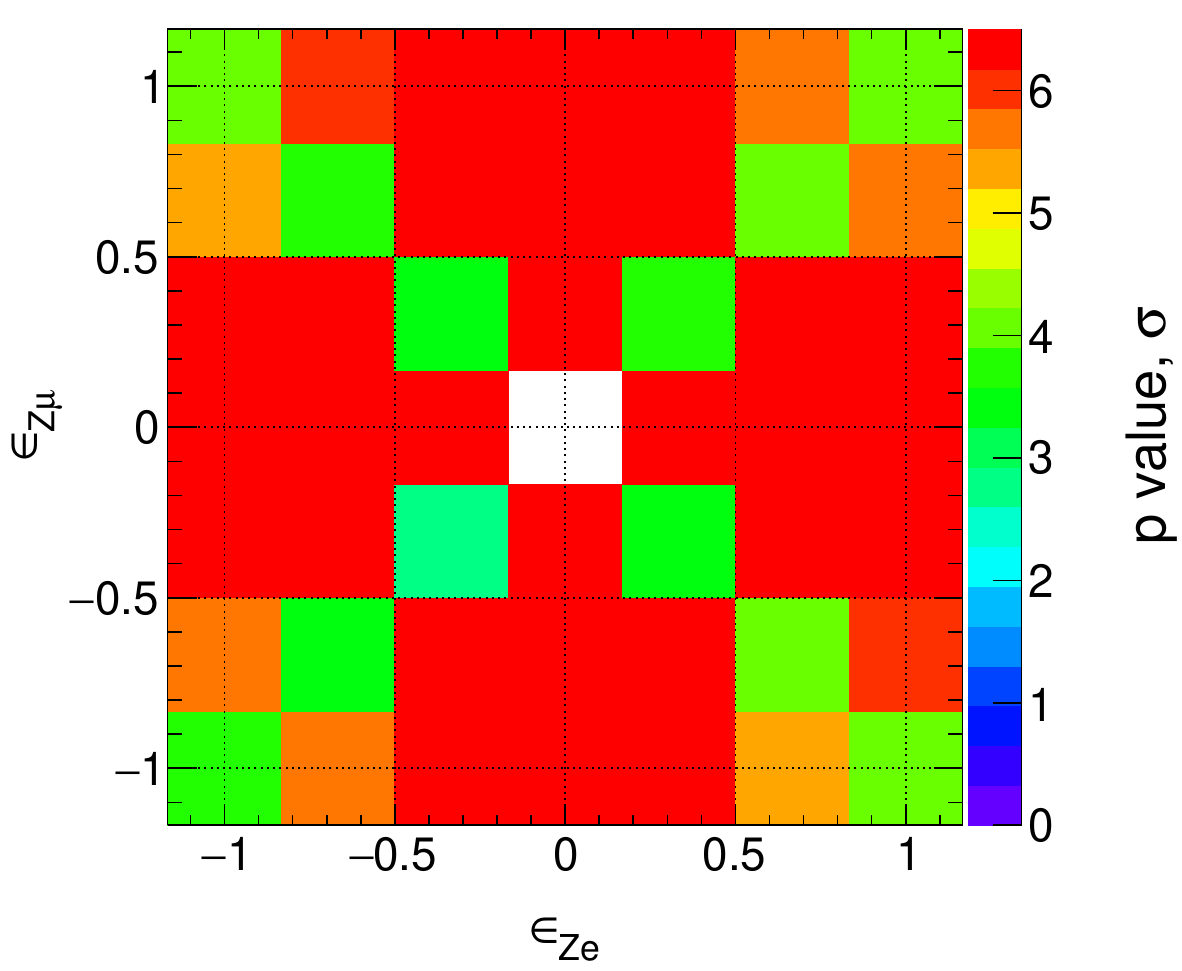}
        \caption{Kinematics and relative normalization}
        \label{fig:scen5b}
    \end{subfigure}

    \caption{
        The left figure is the same as Fig.~\ref{fig:scen4a},
        except that in this case we are considering the benchmark scenario~\ref{scen5}, in which the 
        anomalous PO are
        $\epsilon_{Z e} = \epsilon_{Z e_L} = \epsilon_{Z e_R} $ and
        $\epsilon_{Z \mu} = \epsilon_{Z \mu_L} = \epsilon_{Z \mu_R}$. The right figure is obtained also for scenario~\ref{scen5} but including not only kinematics but also information about the relative normalization of the various $H\to4\ell$ flavor channels.
        \label{fig:scen5}}
\end{figure}

\begin{figure}[t]
    \centering
    \begin{subfigure}[b]{0.48\textwidth}
        \includegraphics[width=\textwidth]{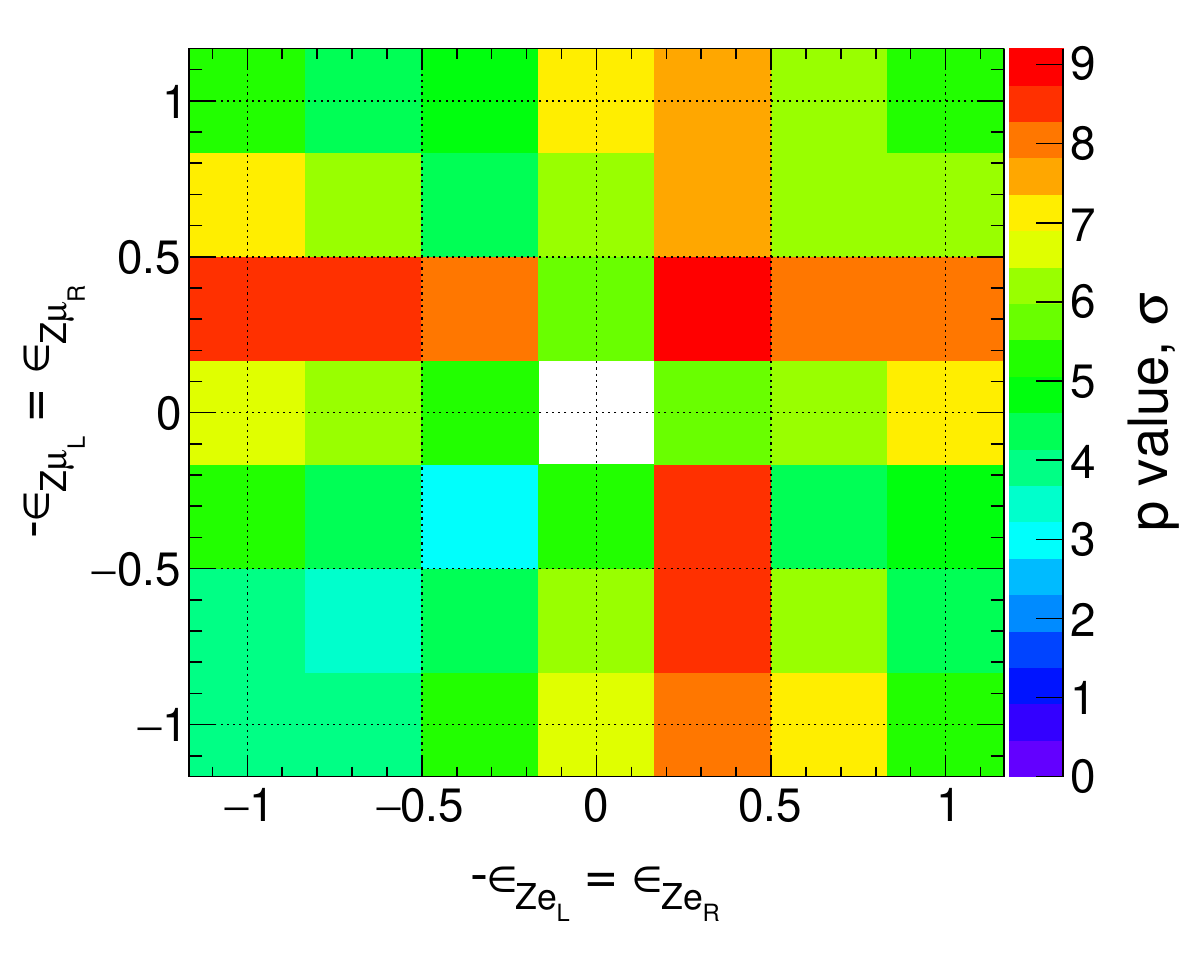}
        \caption{Kinematics}
        \label{fig:scen6a}
    \end{subfigure}
    ~ 
    \begin{subfigure}[b]{0.48\textwidth}
        \includegraphics[width=\textwidth]{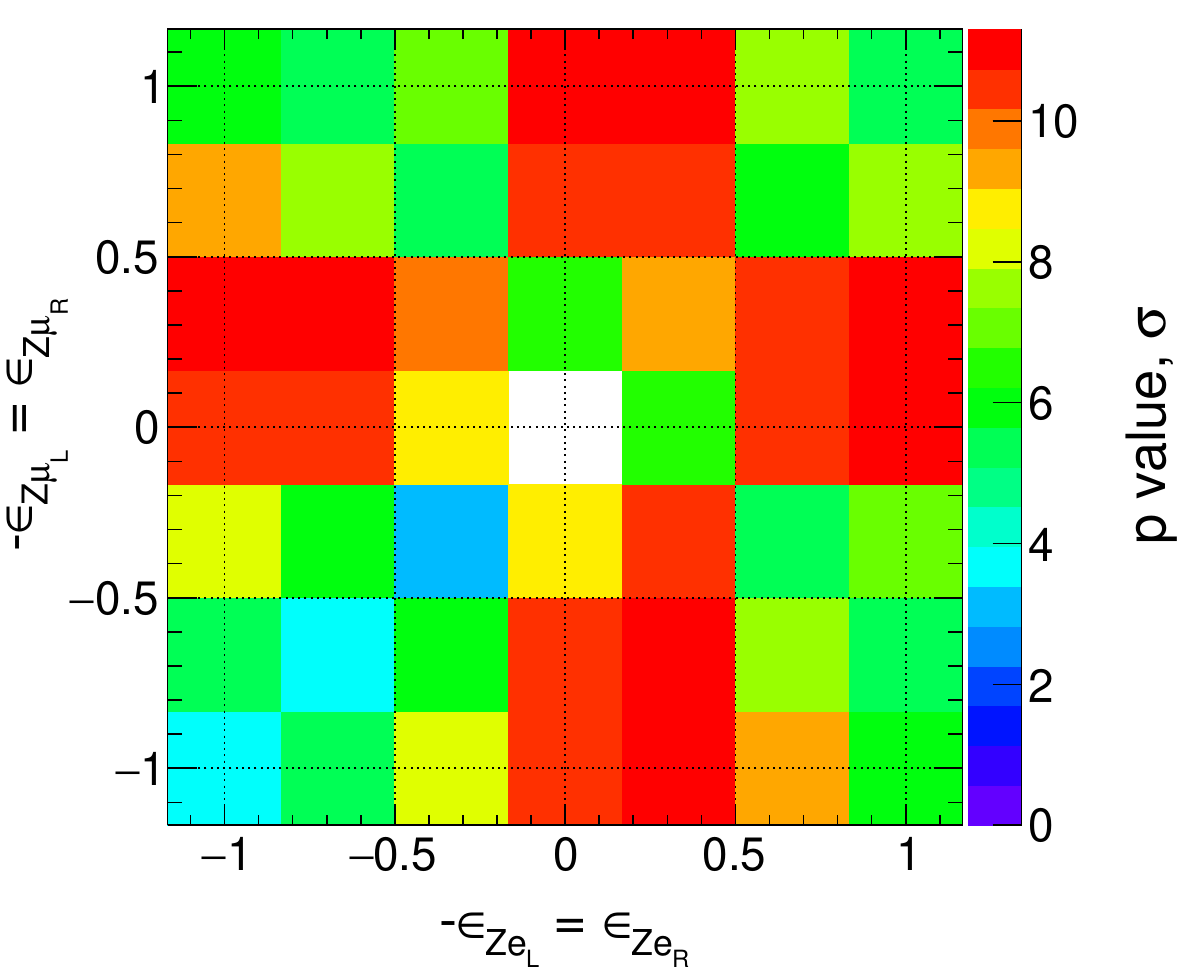}
        \caption{Kinematics and relative normalization}
        \label{fig:scen6b}
    \end{subfigure}

    \caption{
        This figure is the same as Fig.~\ref{fig:scen4a},
        except that here we are considering scenario~\ref{scen6}, in which the
        anomalous PO are
        $\epsilon_{Z e_R} = -\epsilon_{Z e_L} $ and
        $\epsilon_{Z \mu_R} = -\epsilon_{Z \mu_L}$.
        \label{fig:scen6}}
\end{figure}

To ease the comparison with the results of the previous subsection we work again with a fixed luminosity of 100 fb$^{-1}$ 
at 13 TeV. 
Figs.~\ref{fig:scen4}--\ref{fig:scen6} show discovery projections for the non-standard contact terms, $\epsilon_{Z\ell}$, 
present in
scenarios~\ref{scen4}--\ref{scen6}. 
These projections are obtained using the Matrix Element Method (MEM)~~\cite{kondo1, kondo2, kondo3, dalitz,
    oai:arXiv.org:hep-ex/9808029, vigil, canelli, abazov, Gainer:2013iya}, in essentially the same manner as in
Ref.~\cite{Chen:2013waa}. Let us discuss below the technical details.

Since the likelihood ratio, the essential quantity to calculate in the MEM,
reduces to the ratio of squared matrix elements between different coupling 
hypotheses, our analysis is built on such ``discriminants'', specifically
\begin{equation}
    \mathcal{D}_{\rm SM} = \ln \frac{|\mathcal{M}_{\rm H}(p|m_{H} =
        M)|^2}{|\mathcal{M}_{\rm ZZ}|^2},
\end{equation}
and
\begin{equation}
    \mathcal{D}_{\rm exo} = \ln \frac{|\mathcal{M}_{\rm H}(p|m_{H} =
        M)|^2}{|\mathcal{M}_{\rm exo}(p|m_{exo} = M)|^2},
\end{equation}
where $M$ is the invariant mass of the four lepton system,
$|\mathcal{M}_{\rm H}(p|m_{H} = M)|^2$ is the squared matrix element for the
four-lepton momenta, $p$, under the hypothesis that the four leptons are produced 
by the decay of an SM Higgs with mass $M$ (and only tree level couplings),
$|\mathcal{M}_{\rm exo}(p|m_{exo} = M)|^2$ is the squared matrix element for
the four-lepton momenta, $p$, under the hypothesis that the four leptons are 
produced by the decay of a Higgs boson with the exotic (non-SM) PO indicated by 
the axes of the corresponding plot, and $|\mathcal{M}_{\rm ZZ}|^2$ is the squared matrix 
element for the four-lepton momenta $p$ under the hypothesis that the four 
leptons are produced by (leading order) SM $q\bar{q} \to 4\ell$ processes.

To perform the analysis, we obtain the 2-D histograms (``templates'' or, in
statistical language, probability mass functions)
in dimensions of $\mathcal{D}_{\rm SM}$ and $\mathcal{D}_{\rm exo}$ using
parton-level signal and background events. The SM normalization per template
is applied as is appropriate, which is derived using the SM
$H \to 4\ell$ and $q\bar{q} \to 4\ell$ processes.\footnote{
    As mentioned above, we assume throughout this analysis that
    the exotic coupling point has the
    same cross section as the SM, \textit{i.e.}, we do not use the overall
    normalization in our analysis.}
The parton-level histograms were obtained using a background sample of 140
thousand SM $q\bar{q} \to 4\ell$ events and a signal sample which, after 
selection cuts, consists of 560 thousand SM $H \to 4\ell$ events.  The SM Higgs 
and $q\bar{q} \to 4\ell$ events were obtained using 
MG5\_aMC@NLO~\cite{Alwall:2014hca}; to obtain the vast set of samples that model
all the $\gtrsim 150$ coupling points, a matrix-element-based event reweighting
was applied, thus enabling a creation of exotic signal templates from the SM
Higgs events (see, \textit{e.g.}, Ref.~\cite{Gainer:2014bta}).
We conservatively account for mass resolution effects, as in
Ref.~\cite{Chen:2013waa}, by including all events within a ``window'' of $6$ GeV 
around the Higgs boson mass, which we take to be $125$ GeV.
Separate histograms were produced for each four-lepton final-state flavor,
\textit{i.e.}, $4e$, $4\mu$, and $2e2\mu$.  This allows us both to consider the
effects of interference, which arise between the amplitudes for different
lepton pairings in the $4e$ and $4\mu$ final states (see, \textit{e.g.},
Ref.~\cite{Avery:2012um}), and to consider the effects of flavor non-universal
anomalous PO 
in scenarios~\ref{scen5} and ~\ref{scen6}.

As the $\mathcal{D}_{\rm exo}$ discriminant depends on the chosen coupling point, we 
need to obtain templates for the three final states, each of different lepton
flavor, and for the three event types ``SM Higgs", ``exotic PO", and
``background", for all $7 \times 7$ coupling points that we evaluate for each 
scenario: a total of $3 \times 3 \times 7 \times 7 = 441$ templates.  For 
scenario~\ref{scen4}, we construct templates at additional coupling points to make the 
plot of the expanded (``zoomed-in'') region 
(Fig.~\ref{fig:scen4b}). Having the necessary templates, we (a) produce the 
distributions of log-likelihood\footnote{
    The probability mass functions, which are the above-described templates, times the Poisson probability density 
    functions are used to evaluate the likelihoods.}
values for $50000$ $4$-lepton pseudo-experiment events per 
coupling point, with the number of expected signal and background events 
given by the SM Higgs and background event yields, respectively, from
Ref.~\cite{Chen:2013waa} and (b) produce the distributions of log-likelihood values
for $50000$ $4$-lepton pseudo-experiment events per 
coupling point, with the number of expected $4e$, $4\mu$, and $2e2\mu$ 
events in the ratio predicted for the PO point, but with the total number
of $4e$, $4\mu$, and $2e2\mu$ events set to the SM value.

Our immediate aim, for each exotic PO point, and for each of these two analyses,
was to determine the mean
separation power between the ``SM Higgs" and the ``exotic PO". This is done
by evaluating the distance in $\sigma_{SM}$ between the means of corresponding
log-likelihood distributions. Such a separation criterion can be loosely thought
of as a mean p value and is defined here as the following:
\begin{equation}
    \text{p value} = \frac{<\text{exotic PO}> - <\text{SM Higgs}>}{\sigma_{SM}},
    \label{eq:p-value}
\end{equation}
where $<\text{exotic PO}>$ ($<\text{SM Higgs}>$) is the mean of the
log-likelihood distribution when the exotic coupling point (the SM Higgs) is the
true hypothesis.
To be conservative, we included $5\%$ normalization (yield) uncertainties for
each of the several different (independent) sources that are typical to
experiments, namely
(i) the total event count due to theoretical knowledge,
(ii) the total event count due to partonic density mismodeling,
(iii) the total event count due to generic detector inefficiencies,
(iv) the expected number of exotic PO events,
(v) the expected number of the SM Higgs events, and
(vi) the expected number of background events.
These uncertainties effectively smear (widen) the log-likelihood distributions,
thus increasing the $\sigma_{SM}$ denominator in Eq.~(\ref{eq:p-value}).
In the limit of large numbers of expected events, as for the
100~$\rm{fb}^{-1}$ considered in Figs.~\ref{fig:scen4}--\ref{fig:scen6}, this 
quantity corresponds to the $Z$ value\footnote{By ``$Z$ value'' we mean, 
essentially, the number of ``$\sigma$''  describing the statistical significance of a result.} 
of the expected discovery significance 
and can be scaled in a straightforward way to 
obtain predictions for other luminosities.
The results of the analyses of type (a) are shown in the left panels of
Figs.~\ref{fig:scen4}-\ref{fig:scen6}
and depend on only kinematical properties of the events, while the results
of analyses of type (b) are shown in the right panels Figs.~\ref{fig:scen4}-\ref{fig:scen6} 
and depend on both kinematical information and the relative number of events
in each flavor channel that one would obtain from various PO points.

\subsection{Discussion of Results}

The comparison of the results obtained using rates and kinematical information with the results obtained using only kinematical information in our benchmark scenarios shows that the overall normalization of all $H\to4\ell$ channels provides by far the main sensitivity to the contact terms. However it is important to note that kinematic distributions and ratios of rates do provide additional information that allows one to exclude regions of parameter space with reasonably large nonstandard terms that do not affect the overall normalization much due to accidental cancellations. This is why we do not find a perfect ring-shape region in Fig.~\ref{fig:asn} for scenario 4 (the impact on scenario 5 and 6 is actually much larger).

As mentioned before, the large sensitivity of the overall normalization to contact terms comes however at a price: the implicit assumption that other BSM effects affecting Higgs production (or the SM-like pseudo-observable $\kappa_{ZZ}$) are absent. On the other hand, the relative normalization or the kinematic distribution of the events do not require this assumption. Our results in Figs.~\ref{fig:scen4}--\ref{fig:scen6} show that these normalized observables provide a significant sensitivity to the contact terms in a large part of the parameter space for all of the tested scenarios with 100~$\rm{fb}^{-1}$, indicating that these points are discoverable (or excludable) with the current LHC Run 2 data set.

Particularly strong sensitivity in distinguishing SM Higgs couplings from contact terms PO points occurs when the contributions to the amplitude from the contact terms interfere destructively with the SM Higgs amplitude~\cite{Chen:2013waa}, as can be seen in Fig.~\ref{fig:asn} (left) or Fig.~\ref{fig:scen4} (left)
for $\epsilon_{Z\ell_L} \sim - \epsilon_{Z\ell_R} \sim -0.23$. In scenarios 5 and 6, where lepton-flavor universality violation is present, the comparison of the different $H\to4\ell$ channels represents also a very sensitive probe, as long as couplings to electrons and to muons are not too similar. This latter effect is clearly seen comparing the left and right plots in Figs.~\ref{fig:scen5} and \ref{fig:scen6}.

It is easy to note that the separation power for different coupling points obeys certain symmetries: the swapping of electrons and muons (x with y), in Figs.~\ref{fig:scen5} and \ref{fig:scen6}, does not change the sensitivity for the scenarios~\ref{scen5} and \ref{scen6}, while the flip of any coupling sign, in Fig.~\ref{fig:scen5}, gives an approximately symmetric result.

\section{Conclusions}
\label{sec:conclusions}

Studies of the Higgs boson in the golden, four-lepton, final state
remain important in LHC Run 2 and beyond.  A useful toolkit for describing
the probed Higgs-to-diboson (and, more generally, $HZ\ell\ell$) interactions exists 
and contains three major methods for parameterizing couplings. 
These consist in the specification
of the most general $H\to VV$ amplitude compatible with the symmetries assumed (the amplitude approach),
the specification of the most general $H\to VV$ Lagrangian terms consistent with
these same symmetries (the Lagrangian approach), and the use of the most general decomposition of the  
$H\to 4\ell$ on-shell amplitude in terms of pseudo-observables (the PO approach).
The latter can be considered a generalisation of the two other methods (widely adopted so far in the 
experimental analyses) able, in particular, to cover also more general BSM frameworks. 

In this paper we have compared the use of PO to the
other approaches. We showed how to translate between
the conventions used by the ATLAS and CMS experiments and in selected
theoretical papers, and the parameterization of PO for
the four-lepton case specified in Ref.~\cite{Gonzalez-Alonso:2014eva}.
We have also provided
projections for the LHC Run 2 sensitivity to departures from SM
couplings in this channel, 
and we have analyzed the role of kinematic distributions and (ratios of) rates in such $H\to4\ell$ studies. 
In doing so, we have both demonstrated the
use of PO and illustrated their relationship to other
parametrization methods.

Our work here has aimed at expanding the experimental and
phenomenological toolbox for studies of the four-lepton final state.
This channel, which provides a treasure trove of
information about the underlying coupling of the Higgs to bosons (and leptons),
will play an justifiably important role in the LHC physics program for
years to come.
It is our hope
that subsequent studies of the golden channel will lead to a deeper
understanding of the Higgs and, perhaps, help inaugurate a golden age
of BSM discovery in particle physics.

\acknowledgments
JL, AK, KM, PM, GM, SM, and AR thank their CMS colleagues for useful
discussions. Work supported by DOE Grants No.~DE-SC0010296 and
DE-SC0010504; the Swiss National Science Foundation under contract 200021-159720; 
and a Marie Sk\l{}odowska-Curie Individual Fellowship of the European Commission's Horizon 2020 Programme under contract number 745954 Tau-SYNERGIES.
The research at Cornell was supported by National Science Foundation grant PHY-1607126.
MP is supported by the National Research Foundation of Korea (NRF)  grant funded by the Korea government (MSIT) (No. NRF-2018R1C1B6006572).
This document was prepared in part using the resources of the Fermi National Accelerator Laboratory (Fermilab), a U.S. Department of Energy, Office of Science, HEP User Facility. Fermilab is managed by Fermi Research Alliance, LLC (FRA), acting under Contract No. DE-AC02-07CH11359.
This work was performed in part at the Aspen Center for Physics,
which is supported by National Science Foundation grant PHY-1066293.




\end{document}